\documentclass[aps,prl,twocolumn, showpacs, groupedaddress]{revtex4}

\begin{document}
 

\title{Alternate Gauge Electroweak Model}

\author{Bill Dalton}
\affiliation{Department of Physics, Astronomy and Engineering Science,
St Cloud State University
}
\date{ \today}

\begin{abstract}
We describe an alternate gauge electroweak model that permits neutrinos with mass, and at the same time explains why right-handed neutrinos do not appear in weak interactions. This is a local gauge theory involving a space $[V]$ of three scalar functions. The standard  Lagrangian density for the Yang-Mills field part and Higgs doublet remain invariant.  A major change is made in the transformation and corresponding Lagrangian density parts involving the right-handed leptons. A picture involving two types of right-handed leptons emerges.  A dichotomy of matter on the $[V]$ space corresponds to coupled and uncoupled right-handed Leptons. Here, we describe a covariant dipole-mode solution in which the neutral bosons $A_\mu$ and $Z_\mu$ produce precessions on $[V]$. The  $W_\mu^{\pm}$ bosons provide nutations on $[V]$, and consequently, provide transitions between the coupled and uncoupled regions. To elucidate the $[V]$ space matter dichotomy, and to generate the boson masses, we also provide an alternate potential Lagrangian density.

\end{abstract}


\maketitle 
Neutrino mass is implied via neutrino oscillations. Right-handed neutrinos have never been observed in weak interactions. With the standard electroweak model \cite{w}, \cite{S}, it is difficult to reconcile these combined observations. This difficulty has been brought into focus more sharply with the recently reported evidence \cite{sn} of a sterile neutrino.
In \cite{N} an explanation for why right-handed neutrinos do not appear in weak interactions was presented. That work, based on much earlier work \cite{dc} on nonlinear realizations, also permitted sterile right-handed neutrinos. In \cite{N}, a picture involving a dichotomy of matter, with coupled and uncoupled right-handed leptons respectively, emerged. No means of transitions between these matter regions was provided. A key feature of this model was use of transformation eigenstates for the right-handed leptons. 

Here, we use an extended realization of the gauge group $U(1)$ to present a picture similar to \cite{N}.  We use  $UE(1)$ to indicate these extended realizations. Transformation eigenstates for the right-handed leptons are used. Three real scaler functions $V^k$  are used. As in \cite{N} we obtain disjoint regions of matter on the manifold of $[V]$  not connected by the $UE(1)$ realizations. To provide transitions between matter regions, we introduce in this paper, covariant dipole-mode relations. With these, the neutral bosons $A_\mu$ and $Z_\mu$ produce precessions and the $W_\mu^{\pm}$ bosons produce nutations on the $[V]$ manifold. We discuss possible physical significance of this dipole-mode.

The alternate  lepton Lagrangian density presented below is based on these realization acting on left and right stacked spinors  $\bar{\mathbf{L}} = \left(\begin{array}{cc} \bar{\nu_L}&\bar{e_L}\end{array}\right) $ and   $\bar{\mathbf{R}} = \left(\begin{array}{cc} \bar{\nu_R}&\bar{e_R}\end{array}\right) $.  The left component $ \bar{\mathbf{L}}$ is the same as in the standard model. Here, unlike in the standard model, we use a right-handed component  $ \bar{\mathbf{R}}$ with two right-handed leptons. The notation $e$ and $\nu$ are simply used to label the bottom and top leptons. The actual interpretation follows from the interactions. 
The infinitesimal transformations are
\begin{eqnarray}\label{at}
\mathbf{L}^{\prime} =\mathbf{L} +\frac{i}{2}\Lambda(x)[T,\mathbf{L}], \hspace{.3cm}
[T,\mathbf{L}]     =   (Y\mathbf{I}+\mathbf{H)}\mathbf{L}, \hspace{.3cm} \nonumber \\\ 
 \mathbf{R}^{\prime} =\mathbf{R} +\frac{i}{2}\Lambda(x)[T,\mathbf{R}], \hspace{.3cm}
[T,\mathbf{R}]     =   (Y\mathbf{I}+\mathbf{H)}\mathbf{R},\nonumber \\\ \mathbf{H} = \left(\begin{array}{cc}h_3 \mathcal{U} & \sqrt{2}h^-\mathcal{U} \\  \sqrt{2}h^+\mathcal{U} &  -h_3\mathcal{U} \end{array}\right) , h^{\pm} = \frac{1}{\sqrt{2}}(h_1 \pm ih_2).
\end{eqnarray}
The matrices  $\mathbf{I}$ and $U$ are unit vectors in eight and four dimensions respectively. We take $Y = -1$ for both the right-handed and left-handed components. This is another change from the standard model.  The three $h^k(x)$ are unit vectors in the internal $[V(x)]$ space via $ h^k = \frac{V^k}{V}$ where $V^2 = V^kV_k$  and the $V^k$ are group scalars, $(\delta V^k  =0)$.

For the covariant derivative acting on $\mathbf{L}  $ we use the same general form and notation as in the common electroweak gauge model. 
\begin{eqnarray}\label{P}
D_\mu \mathbf{L} = \partial_\mu \mathbf{L}   - \frac{i}{2}\mathbf{P}\mathbf{L},  \hspace{3cm} \nonumber \\\ \mathbf{P} = \left[\begin{array}{cc} (g W^3_\mu-g^\prime \beta_\mu) \mathcal{U} & g (W^1_\mu-i W^2_\mu) \mathcal{U} \\ g (W^1_\mu+i W^2_\mu) \mathcal{U} &( -g W^3_\mu-g^\prime  \beta_\mu) \mathcal{U} \end{array}\right] \hspace{.4cm}  \nonumber \\
=  \left[\begin{array}{cc} N Z_\mu \mathcal{U} & g\sqrt{2}W^- \mathcal{U} \\ g\sqrt{2}W^+ \mathcal{U} &[ -N\cos{2\theta_w} Z_\mu-2qA_\mu] \mathcal{U} \end{array}\right] 
\end{eqnarray}
We use the standard potential relations 
\begin{equation}\label{pr}
\left(\begin{array}{c}W_\mu^3 \\\beta_\mu\end{array}\right)=\left(\begin{array}{cc}cos(\theta_w) & sin(\theta_w) \\-sin(\theta_w) & cos(\theta_w)\end{array}\right) \left(\begin{array}{c}Z_\mu \\A_\mu\end{array}\right),
\end{equation}
with the parameter notation $\cos(\theta_w) = \frac{g}{N}$, $\sin(\theta_w)= \frac{g^{\prime}}{N}$, $N = \sqrt{(g^{\prime})^2 + g^2}$ with the charge $q = \frac{g^{\prime} g}{N}$.
The transformations on the potentials are given by
\begin{eqnarray}\label{Twp}
B_\mu^{\prime} = B_\mu + \delta B_\mu, \hspace{.3cm} \delta B_\mu = \frac{1}{g^{\prime}}\partial_\mu \Lambda,  \nonumber \\\ 
(W^{\prime})_\mu^{l} = W^l_\mu + \delta W^l_\mu, \nonumber \\\
\delta W_\mu^l  =  - \Lambda h^i \epsilon^{ikl} W_\mu^k +\frac{1}{g}\partial_\mu ( \Lambda h^l).
\end{eqnarray}
Notice that the action on the $W^l_\mu$  potentials involves a cross term with the $h^k$ and derivatives of the parameter $\Lambda(x)$ as well as the scalars fields $h^k(x)$. The following (dipole-mode) equations are covariant. 
\begin{eqnarray}\label{dp}
\partial_\mu h^l  =  c g h^i \epsilon^{ikl} W_\mu^k , \hspace{.3cm} \delta W_\mu^l  =  \frac{1}{g}h^l\partial_\mu ( \Lambda).
\end{eqnarray}
With $c=1$, the transformations on the $W_\mu^l$ simplifies to $ \delta W_\mu^l  =  \frac{1}{g}h^l\partial_\mu ( \Lambda)$. With the dipole-mode relations, it is convenient to view the $h^k$ as internal dipole components. 
With  $\mathbf{\Gamma}^\mu= \gamma^\mu\mathbf{I}$ the Lagrangian density for this left-handed lepton has the standard form.
\begin{eqnarray}\label{K1}
\mathcal{L}_L= \frac{1}{2}[i  \tilde{\mathbf{L}}\mathbf{\Gamma}^\mu D_\mu \mathbf{L}  +(i  \tilde{\mathbf{L}}\mathbf{\Gamma}^\mu D_\mu \mathbf{L} )^\ast] 
\end{eqnarray}

Now we come to a major difference between this model and the standard one. To construct an invariant Langrangian density component for the right-handed lepton part,  consider the following eigenstate equation.
\begin{eqnarray}
\mathbf{H}\mathbf{R}=\mathbf{h} \cdot \mathbf{\sigma} \mathbf{R} = \lambda \mathbf{R}, \hspace{.3cm} \lambda=\pm1.
\end{eqnarray}
Because the operator $\mathbf{H}$ appears in the transformation (\ref{at}), this eigenstate relation is covariant. We have two eigenstates. For the $\lambda=-1$ case, the transformation action in (\ref{at}) becomes $ \delta \mathbf{R}^-     =- i\Lambda\mathbf{R}^-$. This is a local diagonal realization, and we have labeled the eigenstate with the eigenvalue. Notice that we get a ($-1$) from the $Y\mathbf{R^-} = -1\mathbf{R^-}$ condition, and a second ($-1$) from the eigenvalue $\lambda=-1$, giving a net factor of (-2).  In the standard gauge picture ( with the notation of \cite{N} ), this factor is obtained by imposing the condition  $Ye_R = -2 e_R$ .  Because the transformation is diagonal, we must use a diagonal covariant derivative.
\begin{eqnarray}\label{D:3}
D_\mu^-\mathbf{R}^-  = \partial_\mu \mathbf{R}^- +  iB_\mu g^{\prime} \mathbf{R}^- \nonumber \\\
g^{\prime}B_\mu =  -N\sin{\theta_w}^2 Z_\mu +qA_\mu 
\end{eqnarray} 
Except that $\mathbf{R}^-$ includes the right-handed neutrino component, this expression has the same form as the covariant derivative for the "singlet"  component of the standard electroweak model. The  eigenvector equations for the $\lambda = -1$ state are.
\begin{eqnarray}\label{C1}
(1+h_3)\nu_R^- + \sqrt{2}h^-e_R^-= 0, \nonumber \\\ \sqrt{2}h^+\nu_R^- + (1-h_3)e_R^- = 0
\end{eqnarray}
These two equations require that the right-handed lepton $\nu_R^-$ vanish at the north pole $h_N= [ h_1=0, h_2=0, h_3=1 ]$, and that $e_R^-$ vanish at the south pole $h_S =[ h_1=0, h_2=0, h_3=-1 ]$. This is a plausible explanation for why right-handed neutrinos do not participate in weak interactions.

For the $\lambda=+1$ eigenstate, the terms in the transformation cancel, so that we have $\delta\mathbf{R}^+     =0$ with $D^+_\mu \mathbf{R}^+ = \partial_\mu\mathbf{R}^+ $. These right-handed leptons have a null transform under these $UE(1)$ realizations, and thus do not have a mandatory potential. Both $e_R^+$ and $\nu_R^+$ are sterile right-handed leptons. They do not participate in electroweak interactions. Neither can represent an electron. The  eigenvector equations for the $\lambda = +1$ state are
\begin{eqnarray}\label{C2}
(-1+h_3)\nu_R^+ + \sqrt{2}h^-e_R^+ = 0, \nonumber \\\ \sqrt{2}h^+\nu_R^+ - (1+h_3)e_R^+ = 0.
\end{eqnarray}
These two equations require that  $e_R^+$ vanish at the north pole $h_N= [ h_1=0, h_2=0, h_3=1 ]$, and that $\nu_R^+$ vanish at the south pole $h_S =[ h_1=0, h_2=0, h_3=-1 ]$. 

We have the following invariant forms.
\begin{eqnarray}\label{KR-}
\mathcal{L}_R^-= \frac{1}{2}[i  \tilde{\mathbf{R}}^-\mathbf{\Gamma}^\mu D_\mu^- \mathbf{R}^-  +(i  \tilde{\mathbf{R}}^-\mathbf{\Gamma}^\mu D_\mu^- \mathbf{R}^- )^\ast] \nonumber \\\  \mathcal{L}^-_m = -m^- [ \tilde{\mathbf{L}}\mathbf{R}^- + \tilde{\mathbf{R}}^-\mathbf{L} ] \hspace{2.7cm} \nonumber \\\ 
 = -m^-[\bar{\nu_R}^-\nu_L +  \bar{\nu_L} \nu_R^-]  -m^-[\bar{e_R}^-e_L +  \bar{e_L} e_R^- ].
\end{eqnarray} 
\begin{eqnarray}\label{KR+}
\mathcal{L}_R^+= \frac{1}{2}[i  \tilde{\mathbf{R}}^+\mathbf{\Gamma}^\mu D_\mu^+ \mathbf{R}^+  +(i  \tilde{\mathbf{R}}^+\mathbf{\Gamma}^\mu D_\mu^+ \mathbf{R}^+ )^\ast] \nonumber \\\ 
 \mathcal{L}^+_m = m^+[ \tilde{\mathbf{L}}\mathbf{R}^+ + \tilde{\mathbf{R}}^+\mathbf{L} ]  \hspace{2.7cm} \nonumber \\\
 -m^+[\bar{\nu_R}^+\nu_L +  \bar{\nu_L} \nu_R^+]  -m^+[\bar{e_R}^+e_L +  \bar{e_L} e_R^+ ].
\end{eqnarray}
Because the forms $[ \tilde{\mathbf{L}}\mathbf{R}^{\pm} + \tilde{\mathbf{R}}^{\pm}\mathbf{L} ]$ are invariant, the functions $m^{\pm}$ must also be invariant. This differs from the commonly used Yukawa form where the invariant forms involve a coupling of three field components.  This does not exclude writing the mass as a product, say $m^{\pm} = G^{\pm}F$, where $G^{\pm}$ are constants and $F$ is an invariant function. One possibility is to set  $F = | \Phi |$ where $\Phi$ is  the Higgs doublet. Another choice is to set $F=V$ where $V$ is the magnitude of the $[V]$ space scalars. This latter possibility is discussed below where we describe an invariant potential Lagrangian density to generate the boson masses.

The combined Lagrangian density involving right-handed leptons is $\mathcal{L}_R = \mathcal{L}_R^+ + \mathcal{L}_R^-  +\mathcal{L}_m^+ + \mathcal{L}_m^-$. At the north pole  $\nu_R^- \to 0$ and  $e_R^+ \to 0$, and at the south pole $e_R^- =0$ and $\nu_R^+=0$.  At the north pole, we have the sterile lepton $\nu_R^+$. At the south pole we have the sterile lepton $e_R^+$. Both sterile leptons can have a mass. In the region between the poles, the top and bottom right-handed leptons are coupled via the constraint relations. This means that in this region these leptons are not free of each other. The constraint equations can be used to eliminate one of the components, but at the cost of putting the $h^k$ components into the kinetic and mass terms..  Before the field equations can be constructed for this region, these constraint relations must in some way be imposed. 

The standard Yang-Mills field \cite{YM} Lagrangian density is invariant under these transformations. We use the notation $\mathcal{L}_F $ to indicate this invariant density. Details for the field tensors in the notation used here are given in \cite{N}. We next  consider an invariant form to generate masses for the boson potentials.  The form for this Lagrangian density with the $c = 1$ dipole-mode described above is 
\begin{eqnarray}\label{KPn}
\mathcal{L}_{P} = \frac{1}{2}\partial_\mu V \partial^\mu V + U(2V^2) \hspace{3cm} \nonumber \\\ +  \frac{1}{2}V^2\big[g^2W_\mu^lW^\mu_l  -2 g g^{\prime}W_\mu^l h_lB^\mu  +(g^{\prime})^2B_\mu B^\mu \big] \nonumber \\\
 = \frac{1}{2}\partial_\mu V \partial^\mu V + U(2V^2) +  \frac{1}{2}V^2\big[2g^2W^+_\mu W_-^\mu \nonumber \\\ +N^2Z_\mu Z^\mu  + 2Nq(1-h_3)W_\mu^3B^\mu   - 2N q h_kW_\mu^kB^\mu \big]. 
\end{eqnarray}
In this expression, the sum on $k$ in the last term is over $[1,2]$ only, and $U(2V^2)$ represents the standard hat potential.   We have used the same potential and parameter relations as in the standard model together with the relation  $V_1^2 +V_2^2 = 2V^+V^-$. 
At the north pole, $V^{\pm} = 0$,  giving
\begin{eqnarray}\label{KKL}
\mathcal{L}_{P}  =  \frac{1}{2}\partial_\mu V \partial^\mu V + U(2V^2) \nonumber \\\ +  \frac{1}{2}V^2\big[2g^2W^+_\mu W_-^\mu +N^2 Z_\mu Z^\mu \big] .
\end{eqnarray}
Writing the scalar magnitude as $V = \frac{1}{2}(\nu_0+\eta)$ we have in the limit $\eta \to 0$ the same mass relations $M_W = \frac{\nu_og}{2}$ and $M_Z = \frac{\nu_o N}{2}$ obtained using the Higgs doublet. 

For other regions on the $V$ manifold, the neutral potentials $A\mu$ and $Z_\mu$ are coupled via the $W_\mu^3B^\mu =\frac{1}{N}q(A^\mu A_\mu-Z^\mu Z_\mu) + \frac{1}{N^2}(g^2-(g^{\prime})^2A_\mu Z^\mu $ factor in the third term of (\ref{KPn}).
In the case where the short lived bosons $W_\mu^{\pm} $ and $Z_\mu$ vanish, the invariant $\mathcal{L}_P$ becomes
\begin{eqnarray}\label{KKA}
\mathcal{L}_{P} =  \frac{1}{2}\partial_\mu V \partial^\mu V + U(2V^2)  + q^2V^2(1- \cos{\sigma})A_\mu A^\mu 
\end{eqnarray}
Here, we have used the polar parameterization $ V_3 = V\cos{\sigma}, V^{\pm} = V\sin{\sigma}\exp{(\pm i \theta)}$ and $W_\mu^3 \to \frac{g^{\prime}}{N}A_\mu$. When $\sigma \to 0$, this invariant reduces to the scalar Lagrangian density
\begin{eqnarray}\label{KKS}
\mathcal{L}_{P} =  \frac{1}{2}\partial_\mu V \partial^\mu V + U(2V^2)  
\end{eqnarray}
When $\sigma$ is not zero, the $A_\mu$ is massive, and consequently, cannot represent light. Recall that we have the constraint relations between the right-handed leptons in the non pole region. With this, it becomes clear that the non  pole region represents a different form of matter. 

The above invariant must be considered with the following auxiliary relation that follows from the dipole mode relations.
\begin{eqnarray}\label{au}
\partial_\mu h^l\partial^\mu h_l = g^2[ W_\mu^kW^\mu_k - (h_iW_\mu^i)h^jW^\mu_j]
\end{eqnarray}
With the dipole-mode, one can have transitions between matter in the north pole and non north pole regions. To see this, consider the dipole-mode relations which we rewrite as 
\begin{eqnarray}
\partial_\mu h^3 = i g [h^+W_\mu^- - h^-W_\mu^+], \nonumber \\\ \partial_\mu h^{\pm} = -i g [h^3W_\mu^{\pm} - h^{\pm}W_\mu^3].
\end{eqnarray}
In the language of magnetic dipole theory, these relations involve both precession and nutation. Presence of the bosons $W_\mu^{\pm}$ produce nutations connecting the north pole and non north pole regions. In the absence of $W_\mu^{\pm}$, the scalar potentials $A_\mu$ and $Z_\mu$ produce precessions via the  $ i g h^{\pm}W_\mu^3 $ term. Nutations depend both on the strength of the nutating field and duration. If the $W_\mu^{\pm}$ fields have sufficient strength and duration to produce a $180^o$ nutation, it is possible to convert one pole to the other. 

In summary, we first remark that we have used the same potential relations (\ref{pr}) used in the standard model. At the north pole point, we obtain a massless $A_\mu$ field corresponding to light. We also have the standard mass ratio  $\frac{M_W}{M_Z}$  using  $\mathcal{L}_P$. The usual lepton potential relations are the same as with the standard model. However we get a sterile right-handed neutrino $\nu_R^+$ at this north pole point. Between the north and south poles, a switch in right-handed neutrinos takes place. The total number of leptons remains constant however.

What is the physical interpretation of the current $j_\mu$ that arises via the transformation (\ref{at})? Some insight can be gained by looking at the conserved current density at the poles.  Following \cite{N} we have
\begin{eqnarray}\label{NPa}
j^\mu_N = \bar{e_L}\gamma^\mu e^-_L + \bar{e_R}^-\gamma^\mu e^-_R - i\big[W_-^{\mu\rho}W^+_\rho- W_+^{\mu\rho}W_\rho^-\big] , \nonumber \\\  j^\mu_S = \bar{e_L}\gamma^\mu e^-_L + \bar{\nu_R}^-\gamma^\mu \nu^-_R +i\big[W_-^{\mu\rho}W^+_\rho- W_+^{\mu\rho}W_\rho^-\big] \hspace{.2cm} 
\end{eqnarray}  
This current density has a form similar to that of the electromagnetic current density. However, the currents at the north and south poles differ in two ways. First, there is a switch between $e_R^- $ and $\nu_R^-$.  There is a change of sign from the potential contribution. This arises from the presence of $h_3$ in (\ref{dp}). Before the field equations and conserved currents can be obtained for non-pole points, the lepton constraint relations (\ref{C1}), (\ref{C2}) and the dipole-mode auxiliary relation (\ref{au}) must be incorporated into the Lagrangian density. 

At non-pole points, the right-handed leptons components in each pair are coupled. This region is not directly associated  with light, because there $A_\mu$ obtains a mass, and is coupled to the $Z_\mu$ boson. Light wise, this region is dark. It represents matter in which leptons and photons as we know them don't exist as free particles. In this sense, it is dark matter, and should affect gravity. Here, we use the name  "coupled-matter" because the right-handed lepton components  as well as the $Z_\mu$ and $ A_\mu$ potentials are coupled. At the north pole one member of each lepton pair vanishes and at the south pole, the other member of each pair vanishes. In between the poles, one component in each pair can be expressed via the constraint equations in terms of the other component in the same pair. In essence, we never have more than two right handed leptons. 

At the south pole we have at first glance a similar lepton structure as at the north pole. There is a major difference however. At the south pole, the analysis with $\mathcal{L}_P$ indicates that both neutral vector bosons that contribute to the interaction are very massive and appear in a coupled form. It follows that leptons in this region cannot  represent an electron for instance. This argument depends on use of the standard potential relations in (\ref{pr}). 

It is incorrect to say that the leptons in the non north pole regions are uncharged.  We have the same coupling form. The only difference is that the $A_\mu$ field becomes massive. One important question is could the massive neutral fields  lead to stable, or perhaps short lived, lepton-potential structures. 

The dipole-mode equations are invariant if we make the replacement $W_\mu^l \to \tilde{W}_\mu^l + \alpha\frac{g^{\prime}}{g}h^l B_\mu$ where $\alpha$ is a scalar function, the $\tilde{W}_\mu^l$ transform exactly like the $W_\mu^l$. We may view the term $ \alpha \frac{g^{\prime}}{g}h^l B_\mu$ as the self boson field of the internal dipole. The diagonal matrix elements  $P_{ij}$ in (\ref{P}) retain the same form if we redefine the $Z_\mu$ and $A_\mu$ potentials. 
\begin{eqnarray}\label{rp}
A_\mu = \sin(\theta_w)\tilde{W}_\mu^3 + \cos(\theta_w)B_\mu  \hspace{2.8cm} \nonumber \\\ +\alpha h^3 \cos(\theta_w)\tan^2(\theta_w)B_\mu, \nonumber \\\  Z_\mu = \cos(\theta_w)\tilde{W}_\mu^3 - \sin(\theta_w)B_\mu +\alpha h^3 \sin(\theta_w)B_\mu 
\end{eqnarray}
This self field offers a possible internal dipole-dipole interaction. If we have two non parallel internal dipole fields  $h_k$ and $h^{\prime}_k$ in the same space time location, they cause precession and nutation on each other. If leptons are associated with each field respectively, this dipole-dipole mechanism offers an indirect means of interaction via the lepton constraint coupling.  An obvious conjecture raised within this model is that this internal dipole-dipole interaction could play a role in neutrino oscillations. The self boson fields produced by each could induce nutations between the north and south poles in each case.

Several final points deserve remarks. The above relations can be viewed as an extension of $U(1)$, or a restriction of  $U(1)\times SU(2)$ where the restriction is that each parameter of $SU(2)$ be parallel to the unit vector $\hat{h}$. The transformation eigenstate condition is then imposed on these restricted transformations for the right-handed leptons. The above analysis using the alternate potential with the dipole-mode, is simpler, but leads to similar results found for the boson masses obtained using the general alternate potential of \cite{N} without the dipole-mode relations. The standard Higgs doublet Lagrangian density is invariant under these realizations. \cite{N} 

We have applied the eigenvalue condition only to the right-handed component $\mathbf{R}$. If we also applied the eigenvalue condition to the left-handed component $\mathbf{L}$, we would obtain different physics. The interactions would all be diagonal (involving only neutral currents potentials). For instance, the combinations  $(\mathbf{L}^-,\mathbf{R}^-)$ , and $ ( \mathbf{L}^+, \mathbf{R}^+)$, lead to respective mass term of form 
\begin{eqnarray}\label{S++}
\mathcal{L}_m^{--} = - \frac{2}{1+h_3} m^{--}[\bar{e}_R^-e_L^- +\bar{e}_L^-e_R^-], \nonumber \\\
\mathcal{L}_m^{++} = - \frac{2}{1+h_3} m^{++}[\bar{\nu}_R^+\nu_L^+ +\bar{\nu}_L^+\nu_R^+]
\end{eqnarray}
The $(++)$ case represents a lepton that has a mass, but has no interaction, (i.e.  a sterile lepton, electroweak wise).  Notice that in each case the effective mass is a minimum at the north pole where $h_3=1$. In these expressions, we have used the eigenvalue constraint relations to eliminate the component that vanishes at the north pole. The constraint relations also introduce the $h^k$ into the kinetic terms which can produce an indirect interaction effect. 

We have a system of coupled leptons that morph into a single free, but different lepton, at the respective poles. This would clearly suggest that if we are to understand ordinary matter, we must also understand dark, or coupled matter, and the transitions between them.
\begin{acknowledgments}
The author would like to think Professor Kevin Haglin for many useful discussions and my wife Anne Dalton and son Lloyd Dalton for helping eliminate many grammatical errors. 
\end{acknowledgments}

\end{document}